# ORIENTATION OF THE LINEAR POLARIZATION PLANE OF Hα RADIATION IN PROMINENCES

Suyunova E.Z., Kim I.S., Osokin A.R.

*Sternberg Astronomical Institute of Lomonosov Moscow State University, Moscow*
*suyunova@sai.msu.ru*

*2D distributions of deviations of the polarization plane from the direction tangential to the solar limb (angle χ) and the sign of χ are presented for Hα prominences of March 29, 2006. The obtained values of χ are in agreement with non-eclipse coronagraphic measurements and indicate the existence of longitudinal magnetic fields. The 2D distributions of the sign of χ show the existence of both «+» and «-» polarities for each prominence. An interpretation in the frame of the existence of oppositely directed magnetic fields is noted.*

## 1. Introduction

So far, magnetic field measurements in prominences are carried out occasionally. The main factors complicating "magnetic" measurements of the linear polarization in prominences are the scattered light in the telescope ($\approx 1.6 \cdot 10^{-5} B_{sun}$ at prominence heights), sky brightness ($\approx 3 \cdot 10^{-5} B_{sun}$ in the optical continuum) and the coronal continuum ($\approx 10^{-6} B_{sun}$), where $B_{sun}$ is the brightness of the solar disk center in 1 A of nearby continuum. During the totality of solar eclipses, the sky brightness and the scattered light in a telescope are negligible, and the main input to the "parasitic" background is the coronal continuum. The increase of the "signal-to-noise" ratio is expected to be 1-2 orders of magnitude [Kim et al, 2011]. High-precision measurements of linear polarization parameters become possible.

Magnetic structure of prominences and, particularly, normal or inverse polarity of their magnetic field is still an open question. It is known that in the absence of magnetic field the polarization plane of the Hα radiation is tangential to the solar limb (angle $\chi = 0$). Therefore, the measurements of angle $\chi$ and the direction of deviations (clockwise or counter-clockwise) can provide information on the strength (together with other data) and the direction of their magnetic fields. Below are the results of determination of the polarization plane orientation for Hα prominences observed during the totality of March 29, 2006 total solar eclipse: Çamyuva, Turkey: 30º34′ E, 36º33′ N, $h_{sun} = 55º$, m = 1.05, duration of totality – 3 m 34 s).

## 2. Method, data reduction

We have developed the high-precision linear polarimetry approach for the coronal continuum radiation [Kim et al, 2013]. The approach was modified to apply to the prominence emission lines radiation to obtain 2D distributions of the polarization degree $p$ and polarization plane orientation (the angle $\chi$ and the sign of $\chi$ of polarization plane deviation from the direction tangential to solar limb) with actual accuracy of 2% for $p$ and 2° for $\chi$. Restrictions on the expected value of polarization degree and on the distance to solar disk center were applied in the software to select the prominences from the general dataset. We refer to the distributions as $p$-, $\chi$- and the sign of $\chi$-images, by analogy with conventional images (2D distributions of I Stokes parameter).

A series of 24 consecutive frames corresponding to full rotation of the polarizer in 5 s, centered at 25 s before the third contact ($T_3$) and obtained through the red KC13 filter (effective width of the optical assembly convolution is 40 nm, transmission at Hα equals 0.6 of the peak transmission), was chosen for analysis. The surrounding coronal continuum background was subtracted. Observations, data reduction and preliminary results for the prominence at position angle of 42° are presented in [Suyunova, Kim, Popov, 2012].

### 3. Results

Fig.1 shows the corrected $\chi$- and sign of $\chi$-images for four prominences at position angles of 42°, 251°, 287° and 307°, obtained with polarization "resolution" of 8″ and illustrates reliability of our polarization degree measurements. The left part shows the theoretical dependence of polarization degree $p$ on height $h$ above solar limb in the absence of magnetic field (the solid line), calculated by us via Zanstra's equations [Zanstra, 1950]. The measurements are shown by color squares for four prominences: blue – P1 42°, green – P2 251°, red – P3 287°, black – P4 307°. Localization of the most part of points below the calculated curve is explained by the depolarizing effect of longitudinal magnetic field in prominences, i.e. the Hanle effect [Bommier et al, 1994]. Error bars are shown. The measurements were made near to $T_3$, therefore most of the E-limb is below the disk of the Moon. Nevertheless, the upper part of the prominence at 42° was observable from $h = 70″$. The central part of Fig.1 shows $\chi$-images of Hα-prominences: three prominences on the W-limb and one on the E-limb. The solar and lunar limbs, prominence position angles, scales for $h$ and $\chi$ are shown. The scale step of 4° is chosen for presentation purposes. The measured values agree with the results of non-eclipse coronagraphic filter linear polarimetry of prominences for the same height interval, taking into account the accuracy of our measurements.

The sign of $\chi$-images are presented on the right. The clockwise deviations of the polarization plane from direction tangential to solar limb are shown with white color and labeled with the «+» sign, counter-clockwise deviations – with black color and the «–» sign. The uniform gray background corresponds to $\chi = 0$. The sign of $\chi$-images reveal the existence of a predominant polarity with inclusions of the opposite polarity in prominences at position angles 42º and 307º. According to the daily solar activity maps of the Kislovodsk Mountain Astronomical Station of Pulkovo observatory they are quiescent prominences. The prominence at 42º was observed in radio by RATAN-600 (eclipse magnitude is 0.998): brightness temperature $T_b$ = 4500-9500 K, electron concentration $n_e = 10^9$ cm$^{-3}$, the estimated magnetic field strength = 100-550 G [Golubchina et al, 2008].

The prominence at P = 287º belongs to the active region filament class and is characterized by the existence of both «+» and «–» polarities as well. The brightness of the prominence at P = 251º was low. That is why the reliable determinations were made in several "points". Anyway, both «+» and «–» polarities exist. We cannot compare our results with other ones because of absence of similar measurements made by others.

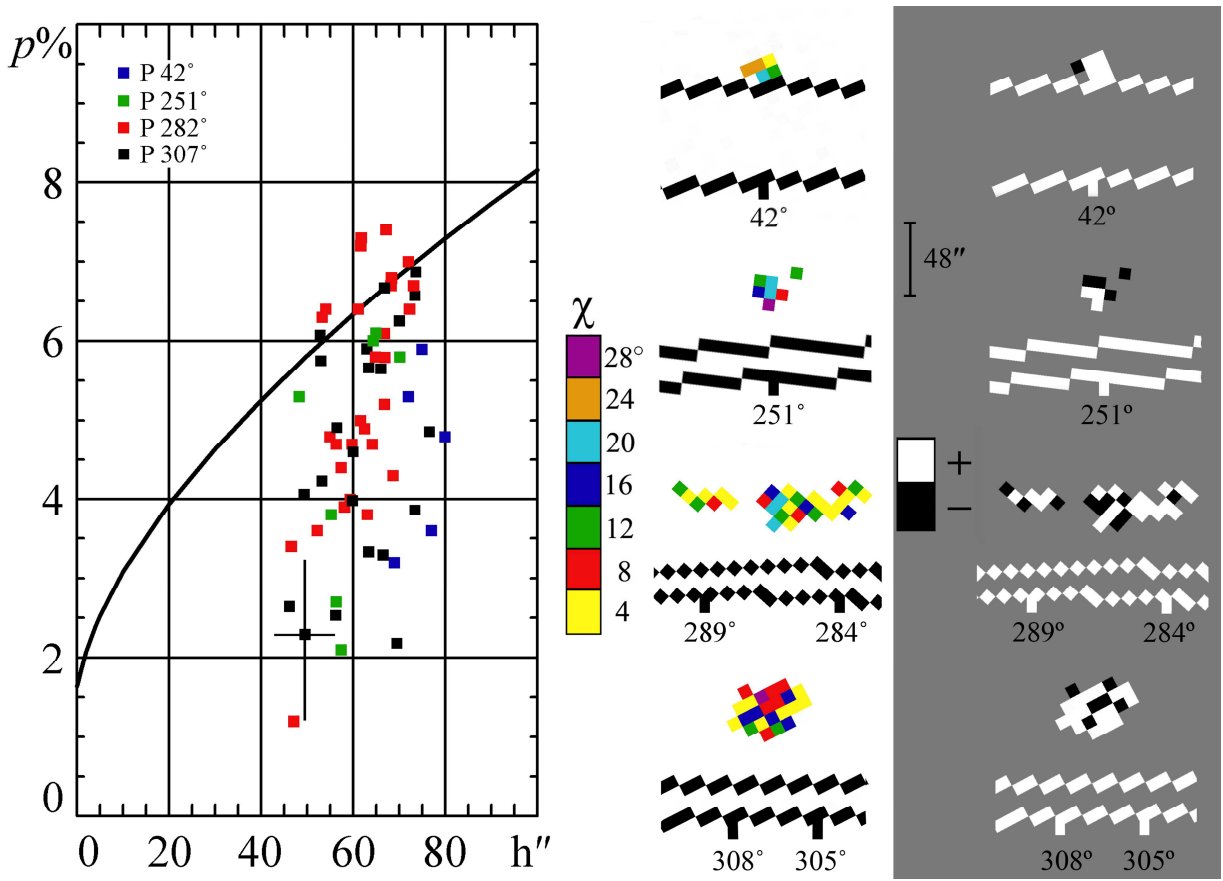

Fig.1. Left: The calculated dependence *p(h)* and measurements. Center: χ-images. Right: sign of χ-images.

**Conclusion.**

Analysis of the sign of χ-images of Hα-prominences of March 29, 2006 reveals the existence of both «+» and «–» polarities in each prominence, which can be explained by the existence of oppositely directed magnetic fields. This excludes the direct classification of these objects as prominences with either normal or inverse polarity.
3

The reported study was partially supported by RFBR, research project No. 14-02-01225.